\documentclass[aps,prl,twocolumn,floatfix,superscriptaddress]{revtex4}

\usepackage{amssymb,amsmath,amstext}                %%   American Physical Society math etc extensions
\usepackage{graphicx}                                               %%   Include figure files                                            %%   Manage links
\usepackage{epstopdf}                                               %%   help with eps -> pdf 
\usepackage{color}                                                     %%   change color
\usepackage{bm}                                                        %%   bold math                                    
\usepackage{appendix}                                              %%   formating appendicies
\usepackage[utf8]{inputenc}

\usepackage{ulem}
\usepackage{latexsym}
\usepackage[colorlinks=true,citecolor=blue,linkcolor=magenta]{hyperref}

\newcommand{\vect}[1]{\mathbf{#1}}

\def\be{\begin{equation}}
\def\ee{\end{equation}}
\def\bea{\begin{eqnarray}}
\def\eea{\end{eqnarray}}
\def\ra{\rangle}
\def\la{\langle}
\def\bi{\begin{itemize}}
\def\ei{\end{itemize}}
\def\ben{\begin{enumerate}}
\def\een{\end{enumerate}}

\definecolor{dgreen} {RGB}{78,138,21}

\begin{document} 

\title{
%Time quasi-crystals: spontaneous emergence from periodically driven systems 
%Spontaneous emergence of discrete time quasi-crystals
Discrete Time Quasi-Crystals
}

\author{Krzysztof Giergiel} 
\affiliation{
Instytut Fizyki imienia Mariana Smoluchowskiego, 
Uniwersytet Jagiello\'nski, ulica Profesora Stanis\l{}awa \L{}ojasiewicza 11, PL-30-348 Krak\'ow, Poland}
\author{Arkadiusz Kuro\'s} 
\affiliation{
Instytut Fizyki imienia Mariana Smoluchowskiego, 
Uniwersytet Jagiello\'nski, ulica Profesora Stanis\l{}awa \L{}ojasiewicza 11, PL-30-348 Krak\'ow, Poland}
\author{Krzysztof Sacha} 
\affiliation{
Instytut Fizyki imienia Mariana Smoluchowskiego, 
Uniwersytet Jagiello\'nski, ulica Profesora Stanis\l{}awa \L{}ojasiewicza 11, PL-30-348 Krak\'ow, Poland}
\affiliation{Mark Kac Complex Systems Research Center, Uniwersytet Jagiello\'nski, ulica Profesora Stanis\l{}awa \L{}ojasiewicza 11, PL-30-348 Krak\'ow, Poland
}

\begin{abstract}
Between space crystals and amorphous materials there exists a third class of aperiodic structures which lack translational symmetry but reveal long-range order. They are dubbed quasi-crystals and their formation, similarly as the formation of space crystals, is related to spontaneous breaking of translational symmetry of  underlying Hamiltonians. Here, we investigate spontaneous emergence of quasi-crystals in periodically driven systems. We consider a quantum many-body system which is driven by a harmonically oscillating force and show that interactions between particles result in spontaneous self-reorganization of the motion of a quantum many-body system and in the formation of a quasi-crystal structure in time. 
\end{abstract}
\date{\today}

\maketitle

Quasi-crystals are related to spatial structures which can not be reproduced by translation of an elementary cell but reveal long-range order \cite{Janot1994,Shechtman1984,Levine1984}. Quasi-crystals are a subject of research in solid state physics but also in  optics \cite{Kohmoto1987,Steurer2007,Albuquerque2003,Vardeny2013} and ultra-cold atomic gases \cite{Viebahn2018}.

Recently research of crystalline structures has migrated to the time domain \cite{Sacha2017rev} (for phase-space crystals see \cite{Guo2013,Guo2016,Guo2016a,Liang2017}). Indeed, a quantum many-body system can spontaneously self-organize its motion and start moving periodically forming a crystalline structure in the time domain. While the first idea of such time crystals turned out to be impossible for the realization \cite{Wilczek2012,Bruno2013b,Watanabe2015,Syrwid2017} another type of spontaneous formation of crystalline structures in time was proposed. These are the so-called discrete time crystals that are periodically driven quantum many-body systems which break spontaneously discrete time translation symmetry of Hamiltonians and start moving with a period different from the driving period \cite{Sacha2015,Khemani16,ElseFTC,Yao2017}. Discrete time crystals have been already realized in laboratories \cite{Zhang2017,Choi2017,Pal2018,Rovny2018,Rovny2018a} and they draw considerable attention in the literature \cite{Nayak2017,sacha15a,delande17,Lazarides2017,Russomanno2017,Zeng2017,Nakatsugawa2017,Ho2017,
Huang2017,Gong2017,Wang2017,Iemini2017,Mierzejewski2017,Giergiel2018,Bomantara2018,Kosior2018,
Mizuta2018,Giergiel2018a,Tucker2018,Kosior2018a,Yu2018} (see also \cite{Shapere2012,Ghosh2014,Yao2018,Das2018,Alvarez2017,Aviles2017} for classical version of time crystals). In the field of time crystals, quasi-crystal structures have been investigated in classical systems \cite{Flicker2018}, quantum systems \cite{Li2012,Huang2017a,Giergiel2018} and in an experiment on magnon condensation \cite{Autti2018}. In Refs.~\cite{Dumitrescu2018,Peng2018} quasi-crystal response of systems which are driven quasi-periodically in time was demonstrated. Quasi-periodic response of a periodically driven many-body system was analyzed in Ref.~\cite{Luitz2018} but with no spontaneous time translational symmetry breaking process involved.

In the present letter we analyze how a quasi-crystal structure forms due to spontaneous breaking of discrete time translation symmetry of a many-body time-periodic Hamiltonian. 

%%%%%%%%%%%%%%            
\begin{figure} 	            
\includegraphics[width=0.8\columnwidth]{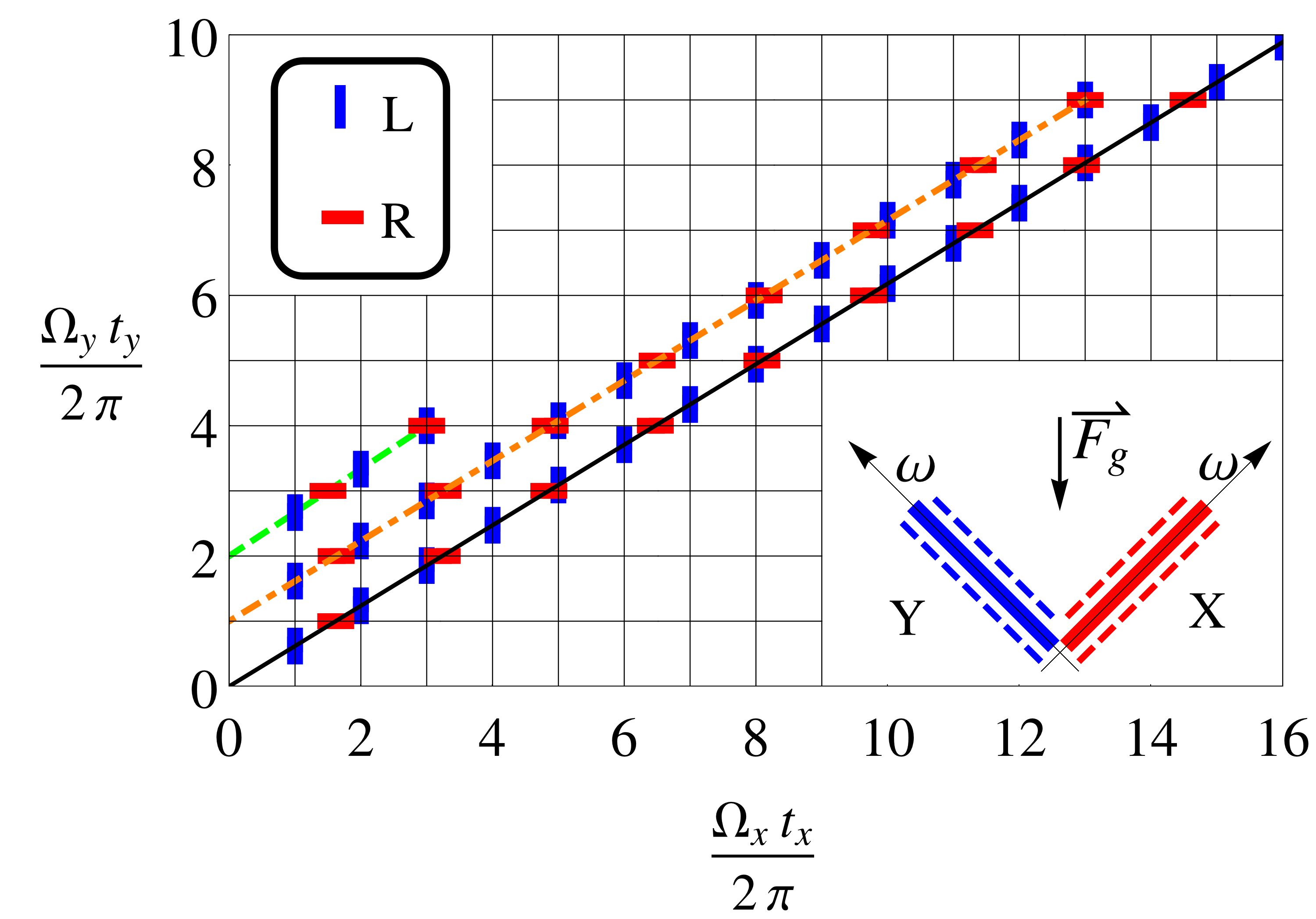}       
\caption{Generation of the one-dimensional Fibonacci quasi-crystal. The solid black line cuts the square lattice. The tangent of the angle that the line forms with the vertical axis is equal to the golden ratio. Consecutive cuts of the line with the vertical ($L$) and horizontal ($R$) lines of the lattice form the Fibonacci quasi-crystal sequence, $LRLLRLRL\dots$. Dashed green and dotted-dashed orange lines correspond to rational approximation of the golden ratio, 3/2 and 13/8, respectively. In the case of ultra-cold atoms bouncing between two mirrors which oscillate with frequency $\omega$ (see the schematic plot in the inset where $\vec F_g$ denotes the gravitational force), the green and orange lines are related to spontaneous formation of finite fragments of the Fibonacci quasi-crystal in the time domain with $\Omega_x/\Omega_y=s_y/s_x=3/2$ and $13/8$, respectively. That is, bounces of atoms off the left $L$ and right $R$ mirrors form a sequence of events that reproduces a fragment of the Fibonacci quasi-crystal of length $s_x+s_y$. Parameters $\Omega_x$ and $\Omega_y$ are frequencies of bouncing of atoms off the left and right mirrors, respectively, see text. The axes of the main figure can be considered as two independent time axes \cite{Flicker2018}, $t_{x,y}$, related to periodic motion along the $x$ and $y$ directions.}
\label{fibonacci}   
\end{figure} 
%%%%%%%%%%%%%% 

One-dimensional (1D) quasi-crystal sequence can be generated by a cut of a square lattice with the help of a line whose gradient is an irrational number \cite{Valsakumar1986,Bombieri1986,Lin1995}. For the Fibonacci quasi-crystal the gradient is the golden ratio and the successive cuts of vertical and horizontal lines of the square lattice produce a sequence $LRLLRLRL\dots$ of two elementary cells which we denote by $L$ and $R$, see Fig.~\ref{fibonacci}. The sequence corresponds to a quasi-crystal structure where there is no translation symmetry but two elementary cells are not distributed randomly so that the sequence reveals long-range order \cite{Janot1994}. A finite fragment of the Fibonacci quasi-crystal sequence can be obtained by cutting the square lattice with a line whose gradient is a rational number that approximates the golden ratio, see Fig.~\ref{fibonacci}. In the following we show how any finite fragment of the Fibonacci quasi-crystal structure can spontaneously emerge in the time evolution of a periodically driven many-body system if interactions between particles are sufficiently strong.

We focus on ultra-cold atoms bouncing between two orthogonal harmonically oscillating mirrors in a 2D model. Such a system can be realized experimentally \cite{Steane95} (for the stationary mirror experiments see \cite{Roach1995,Sidorov1996,Westbrook1998,Lau1999,Bongs1999,Sidorov2002,Fiutowski2013,Kawalec2014}). A single atom bouncing between the mirrors is described, in the frame oscillating with the mirrors
\footnote{In order to switch from the laboratory frame (where one mirror oscillates like $-\frac{\lambda_x}{\omega^2}\cos(\omega t+\Delta\phi)$ along the $x$ direction and the other like $-\frac{\lambda_y}{\omega^2}\cos\omega t$ along the $y$ direction) to the coordinate frame where the mirrors do not move, the unitary transformation $U_{y}=e^{iy\frac{\lambda_y}{\omega}\sin\omega t}e^{ip_y\frac{\lambda_y}{\omega^2}\cos\omega t}$, and a similar one for the motion along $x$, has been applied. We use the gravitational units but assume that the gravitational 
 acceleration is given by $g/{\sqrt{2}}$.} \cite{Sup1q}, by the Hamiltonian \cite{Buchleitner2002}
\be
H=\frac{p_x^2+p_y^2}{2}+x+y+\lambda_x x \cos(\omega t+\Delta \phi)+\lambda_y y\cos(\omega t),
\label{h}
\ee
where $\omega$ is the frequency of the mirrors' oscillations, $\Delta \phi$ the relative phase and $\lambda_{x,y}$ the amplitudes of the oscillations.
The mirrors are located at $x=0$ and at $y=0$ and the gravitational force $\vec F_g$ points in the $-(\vect{x}+\vect{y})$ direction, see inset of Fig.~\ref{fibonacci}. We assume that in the many-body case, $N$ bosons are bouncing between the mirrors and interact via Dirac-delta potential $g_0\delta(\vect{r})$ \cite{Chin2010}. Such contact interactions are determined by the s-wave scattering length of atoms which is assumed to be negative $g_0<0$. The system is periodically driven, thus, we may look for a kind of stationary states which evolve periodically in time. They are eigenstates of the Floquet Hamiltonian $\hat{\cal H}(t)=\hat H-i\partial_t$, where $\hat H$ is a many-body version of (\ref{h}) with the contact interactions between particles included, see \cite{Sup1q}. 
The corresponding eigenvalues are called quasi-energies of the system \cite{Shirley1965,Buchleitner2002}. The discrete time translation symmetry of the time-periodic Hamiltonian, $\hat{\cal H}(t+2\pi/\omega)=\hat{\cal H}(t)$, implies that all Floquet eigenstates must evolve with the driving period $2\pi/\omega$. In the following we show that in the limit when the number of particles $N\rightarrow\infty$ but $g_0N=$const. \cite{Lieb2000}, there are subspaces of the Hilbert space of the system where low-lying quasi-energy eigenstates are fragile because they form macroscopic superposition. Consequently even an infinitesimally weak perturbation, e.g. a measurement of a position of one atom, is sufficient to induce collapse of the many-body state to one of the superimposed states. It results in breaking of the discrete time translation symmetry of the Hamiltonian \cite{Sacha2015}. Interestingly an evolving symmetry broken state can reveal a sequence of events (bounces of atoms off the left $L$ and right $R$ mirrors) which forms a finite fragment of the Fibonacci quasi-crystal in time.

%%%%%%%%%%%%%%            
\begin{figure} 	            
\includegraphics[width=0.8\columnwidth]{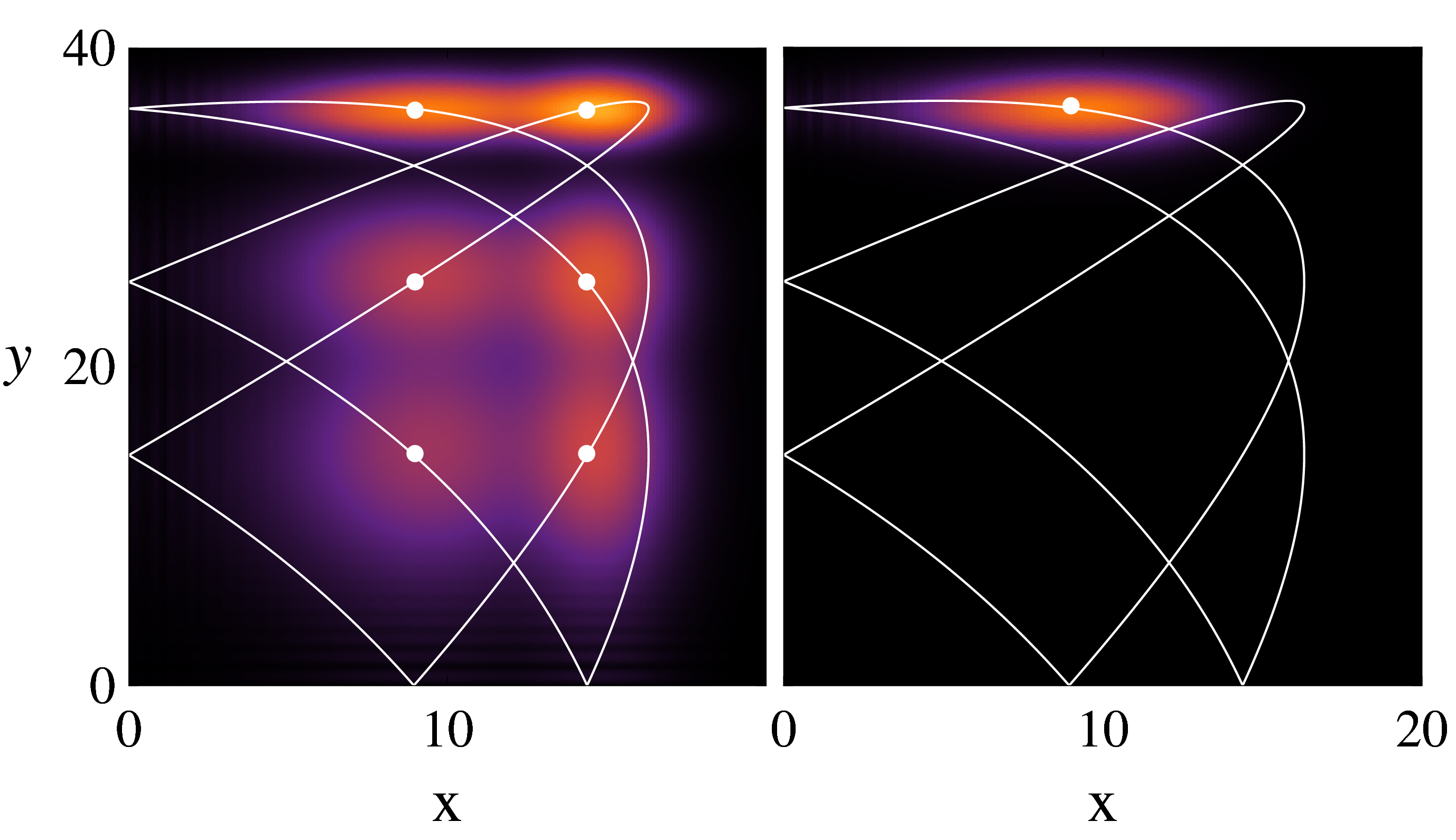}       
\caption{Density of atoms bouncing between two orthogonal oscillating mirrors at $t=2\pi/{3\omega}$. The left ($L$) mirror is located at $x=0$ and the right ($R$) mirror at $y=0$ and the gravitational force $\vec F_g$ points in the $-(\vect x+\vect y)$ direction, see inset of Fig.~\ref{fibonacci}. Left panel is related to the symmetry preserving state which evolves periodically with the driving period $2\pi/\omega$ --- the left and right mirrors are visited by atoms alternately: $LRLRLR$. The presented density consists of $s_xs_y$ localized Wannier-like wavepackets ($s_x=2$, $s_y=3$). The trajectory the Wannier wavepackets are moving along is drawn in the panels. Right panel corresponds to a symmetry broken state where interactions between atoms result in spontaneous breaking of discrete time translation symmetry of the Hamiltonian and emergence of a quasi-crystal structure in time. Atoms are visiting the left and right mirrors in an order that matches the sequence $LRLLR$ i.e. a finite fragment of the Fibonacci quasi-cristal is reproduced because the golden ratio gradient of the line in Fig.~\ref{fibonacci} is approximated by the rational number $\Omega_x/\Omega_y=s_y/s_x=3/2$. The parameters of the system are: $\lambda_x=0.094$, $\lambda_y=0.030$, $\omega=1.1$, $\Delta \phi=2\pi/3$, $g_0N=0$ (left) and $g_0N=-0.022$ (right). The latter results in $U_{\vect i\vect i}N/J=-81$, with $J=4.8\times 10^{-6}$, in the Hamiltonian (\ref{bhh}) that describes an effective $s_x\times s_y$ lattice.
The results are obtained within the quantum secular approach \cite{Berman1977}.
}
\label{trajectory}   
\end{figure} 
%%%%%%%%%%%%%% 

Let us start with the single-particle problem (\ref{h}) which consists of the independent motion along $x$ and $y$ directions. We are interested in a resonant driving of the system, i.e. the frequencies $\Omega_x$ and $\Omega_y$ of the unperturbed particle motion along the $x$ and $y$ directions fulfill $s_x\Omega_x=\omega$ and $s_y\Omega_y=\omega$ with integer $s_{x}$ and $s_y$. The description of a resonantly driven particle can be reduced to an effective tight-binding Hamiltonian \cite{Lichtenberg1992,Buchleitner2002,sacha15a,sacha16,Sup1q}. 
When we switch from a single particle to many bosons resonantly driven, the single-particle tight-binding Hamiltonian becomes the Bose-Hubbard Hamiltonian,
\be
\hat {\cal H}_{\rm eff}=-\frac{1}{2}\sum_{\langle \vect{i}, \vect{j}\rangle}J_{\vect i \vect j}\;\hat a_{\vect i}^\dagger\hat a_{\vect j}+\frac12\sum_{\vect i,\vect j}U_{\vect i\vect j}\;\hat a_{\vect i}^\dagger\hat a_{\vect j}^\dagger\hat a_{\vect j}\hat a_{\vect i},
\label{bhh}
\ee
which is the many-body Floquet Hamiltonian written in a basis of time-periodic functions $W_{\vect{i}=(i_x,i_y)}(\vect{r},t)=w_{i_x}(x,t)w_{i_y}(y,t)$ which are localized wavepackets propagating along the classical resonant orbit with the period $T=s_xs_y2\pi/\omega$ and which play a role of Wannier functions known in condensed matter physics  \cite{Dutta2015}, see Fig.~\ref{trajectory}.
 There are $s_xs_y$ Wannier functions $W_{\vect i}$ which are products of localized wavepackets $w_{i_x}(x,t)$ and $w_{i_y}(y,t)$ moving along the $x$ and $y$ directions with the periods $2\pi/\Omega_x$ and $2\pi/\Omega_y$, respectively. In (\ref{bhh}), $\hat a_{\vect{i}}$'s are the standard bosonic annihilation operators, the nearest neighbor tunneling amplitudes $J_{\vect i \vect j}=-(2/T)\int_0^Tdt\int d^2\vect r W_{\vect i}^*(\vect r,t)[H-i\partial_t]W_{\vect j}(\vect r,t)$ and the coefficients of the effective interactions $U_{\vect i \vect j}=(2/T)\int_0^Tdt\int d^2\vect r g_0|W_{\vect i}|^2|W_{\vect j}|^2$ for $\vect i\ne \vect j$ and similar $U_{\vect i\vect i}$ but by factor two smaller \cite{sacha15a,Giergiel2018,Sup1q}. In the present Letter we choose the amplitudes of the mirrors' oscillations, $\lambda_x$ and $\lambda_y$, so that the resulting amplitudes for nearest neighbor  tunnelings along the $x$ and $y$ directions are the same, $J\equiv J_{\vect i\vect j}$. Typically, the coefficient for the on-site interactions $U_{\vect i\vect i}$ is at least an order of magnitude larger than $U_{\vect i\vect j}$ for long-range interactions ($\vect i\ne\vect j$). 

To conclude this part, the description of the resonantly driven many-body system is reduced, in the time-periodic basis $W_{\vect i}(\vect r,t)$, to the Bose-Hubbard Hamiltonian (\ref{bhh}) \cite{Sup1q,Lacki2013}. The resonant driving is related to non-linear classical resonances where a particle cannot absorb unlimited  amount of energy because transfer of the energy changes a period of motion of the system, a particle goes out of the resonance and the transfer stops \cite{Sup1q}.

For negligible interactions between particles the ground state of $\hat {\cal H}_{\rm eff}$ is a Bose-Einstein condensate $\Psi_0(\vect r_1,\dots,\vect r_N,t)=\prod_{j=1}^N\psi(\vect r_j,t)$, i.e. all atoms occupy a condensate wavefunction $\psi(\vect r,t)\propto \sum_{\vect i}W_{\vect i}(\vect r,t)$ which evolves with the driving period $2\pi/\omega$. Indeed, despite the fact that each $W_{\vect i}$ evolves with the period $T=s_xs_y2\pi/\omega$, after each period $2\pi/\omega$, the Wannier wavefunctions $W_{\vect i}$ exchange their positions so that the condensate wavefunction $\psi(\vect r,t)$ propagates with the driving period, see Fig.~\ref{trajectory}. When the interactions between atoms are attractive and sufficiently strong it is energetically favourable to group all atoms in a single localized wavepacket $W_{\vect i}(\vect r,t)$ \cite{Sacha2015}.
Then, we expect the ground state of $\hat {\cal H}_{\rm eff}$ to be of the form $\Psi_0=\prod_{j=1}^NW_{\vect i}(\vect r_j,t)$ where $\vect i$ can be arbitrary. However, such a state cannot be a Floquet eigenstate of the system because it evolves with the period $T=s_xs_y2\pi/\omega$ while the discrete time translation symmetry of the Hamiltonian requires that all Floquet eigenstates must evolve with the period of the driving $2\pi/\omega$. In order to reconcile the energy and symmetry requirements, the ground state of $\hat{\cal H}_{\rm eff}$ takes the form $\Psi_0\propto\sum_{\vect i}\prod_{j=1}^NW_{\vect i}(\vect r_j,t)$ which is macroscopic superposition of Bose-Einstein condensates \cite{Zin2008,Oles2010}.
However, such a macroscopic superposition is extremely fragile and it is sufficient, e.g., to measure position of one atom and the ground state collapses to one of the Bose-Einstein condensates which form the macroscopic superposition, $\Psi_0\rightarrow \Psi\approx \prod_{j=1}^NW_{\vect i}(\vect r_j,t)$ \cite{Mahmud2002,Oles2010} --- which $W_{\vect i}$ is chosen depends on a result of the measurement. In the limit when $N\rightarrow\infty$ but $U_{\vect i\vect i}N=$const., the latter state is robust and evolves with the period $T=s_xs_y2\pi/\omega$ and thus breaks time translation symmetry of the many-body Hamiltonian \cite{Sacha2015}. The described scenario is an example of a process of spontaneous breaking of time translation symmetry in the quantum many-body system. Similar spontaneous symmetry breaking phenomenon is not present in ~\cite{Luitz2018} because Floquet states are related to single Fock states in the position representation.

%%%%%%%%%%%%%%            
\begin{figure*} 	            
\includegraphics[width=1.6\columnwidth]{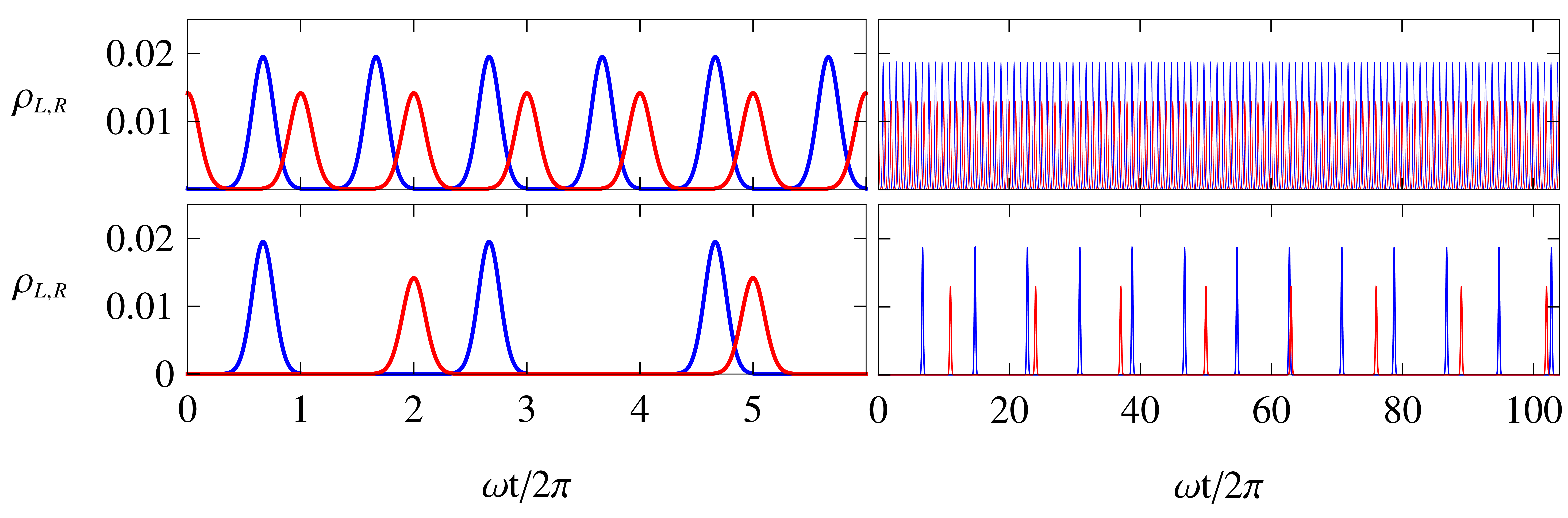}       
\caption{Scaled probabilities for the detection of atoms close to the left mirror $\rho_L(t)$ (blue lines) and close to the right mirror $\rho_R(t)$ (red lines), where $\rho_L(t)=\int dy |\psi(x\approx 0,y,t)|^2$ and there is an analogous expression for $\rho_R(t)$. Top panels are related to symmetry preserving states while bottom panels to states where the discrete time translation symmetry of the Hamiltonian is spontaneously broken. Left panels correspond to $\Omega_x/\Omega_y=s_y/s_x=3/2$ while in the right panels such ratios are equal 13/8. Symmetry preserving states form periodic sequences of the elementary cells  $L$ and $R$ associated with the alternate appearance of maxima of $\rho_L(t)$ and $\rho_R(t)$. In the symmetry broken case, bounces of atoms off the left and right mirrors form a sequence of events that reproduces a finite fragment (of length $s_x+s_y$) of the Fibonacci quasi-crystal which is repeated in the time evolution of the system with the period $T=s_xs_y2\pi/\omega$. The results shown in the left panels correspond to the same parameters as in Fig.~\ref{trajectory}, while in the right panels: $\lambda_x=0.087$, $\lambda_y=0.026$, $\omega=1.77$, $\Delta \phi=\pi/2$, $s_y=13$, $s_x=8$, $g_0N=0$ (top right panel) and $g_0N=-0.029$ (bottom right panel). The latter results in $U_{\vect i\vect i}N/J=-80$ and $J=2.3\times 10^{-6}$ in the Bose-Hubbard Hamiltonian that describes an effective $s_x\times s_y$ lattice.}
\label{LRstructure}   
\end{figure*} 
%%%%%%%%%%%%%% 

In order to describe the system we apply the mean-field approach \cite{Sacha2015,Pethick2002,Giergiel2018a,Sup1q}. The mean-field approximation is valid because the ground state of (\ref{bhh}) for negligible interactions and also symmetry broken states, $\Psi\approx \prod_{j=1}^NW_{\vect i}(\vect r_j,t)$, in the regime of the quasi-crystal formation are Bose-Einstein condensates. The mean-field energy of the system per particle reads $E=-(J/2)\sum_{\langle \vect{i}, \vect{j}\rangle}a_{\vect i}^*a_{\vect j}+(N/2)\sum_{\vect i\vect j}U_{\vect i\vect j}|a_{\vect i}|^2|a_{\vect j}|^2$ \cite{Giergiel2018a,Sup1q} and we are looking for a condensate wavefunction $\psi(\vect r,t)=\sum_{\vect i}a_{\vect i}W_{\vect i}(\vect r,t)$ which minimizes $E$ \cite{Lieb2000,Pethick2002}. In the left panel of Fig.~\ref{trajectory} we show such a wavefunction $\psi(\vect r,t)$ obtained for negligible interactions and for $\Omega_x=\omega/2$ and $\Omega_y=\omega/3$ (i.e. $s_x=2$, $s_y=3$). The wavefunction $\psi(\vect r,t)$ is a uniform superposition of $s_xs_y=6$ localized Wannier wavepackets, it evolves with the period $2\pi/\omega$ and describes atoms bouncing alternately off the left ($L$) and right ($R$) mirrors. If we plot probabilities for the measurement of atoms close to the left mirror, $\rho_L(t)=\int dy |\psi(x\approx 0,y,t)|^2$, and close to the right mirror, $\rho_R(t)=\int dx |\psi(x,y\approx 0,t)|^2$, we can see that maximal values of $\rho_{L,R}(t)$ appear alternately and form a periodic sequence of events $LRLR\dots$, see Fig.~\ref{LRstructure}. 
However, if the interactions are sufficiently strong, i.e. $U_{\vect i\vect i}N/J\lesssim-6.5$, the system chooses spontaneously motion with the period $T=s_xs_y2\pi/\omega$. That is, the mean-field approach shows that the ground state energy is degenerate and the corresponding wavefunctions are not uniform superposition of $W_{\vect i}$. The system prepared in a lowest energy mean-field state breaks discrete time translation symmetry of the many-body Hamiltonian because it evolves with the period different from the driving period. 
For $U_{\vect i\vect i}N/J\lesssim-25$ the symmetry broken degenerate ground states reduce to $\psi(\vect r,t)\approx W_{\vect i}(r,t)$ with accuracy better than 99\% --- which $W_{\vect i}$ is chosen by the system is determined in a spontaneous symmetry breaking process. In Fig.~\ref{trajectory} we show an example of such a ground state wavefunction $\psi(\vect r,t)$ where a single localized wavepacket bouncing between the mirrors is visible. The corresponding probabilities $\rho_{L,R}(t)$ form a sequence of events $LRLLR$, whose length is $s_x+s_y=5$, which is repeated with the period $T$, see Fig.~\ref{LRstructure}. The sequence is a fragment of the Fibonacci quasi-crystal. The time quasi-crystal states predicted by the mean-field approach lives forever. The predictions are valid in the limit when $N\rightarrow\infty$ but $g_0N=$const because then the corresponding symmetry preserving eigenstates of the quantum many-body model (\ref{bhh}) are degenerate and their superpositions, that form the symmetry-broken states, do not decay \cite{Sacha2015}.

It becomes now clear how to realize conditions where any finite fragment of the Fibonacci quasi-crystal emerges due to spontaneous breaking of discrete time translation symmetry of the Hamiltonian: (i) One has to choose a rational number $s_y/s_x$ which approximates the golden ratio and reproduces a given fragment of the Fibonacci quasi-crystal sequence when it is taken as the gradient of the line in Fig.~\ref{fibonacci}. (ii) Then, we know which resonant subspace of the periodically driven many-body system is able to realize such a quasi-crystal, i.e. the subspace corresponding to the frequencies of unperturbed single-particle motion $\Omega_x=\omega/s_x$ and $\Omega_y=\omega/s_y$. (iii) If the many-body system is prepared in a low-lying eigenstate within this subspace, then either atoms are bouncing off the left and right mirrors in the alternate way (if the interactions are negligible) or the bounces on the mirrors form a sequence of events that reproduces a finite fragment of the Fibonacci quasi-crystal (if the interactions are sufficiently strong). In the right panels of Fig.~\ref{LRstructure} we illustrate these two situations for $s_x=8$ and $s_y=13$. In the symmetry preserving case, the probabilities for detection atoms close to the left and right mirrors, $\rho_{L,R}(t)$, show a periodic sequence of maxima $LRLR\dots$. However, when the attractive interactions are sufficiently strong, the discrete time translation symmetry is spontaneously broken and the Fibonacci quasi-crystal $LRLLRLRL\dots$  is formed \cite{Flicker2018}. We would like to stress that the quasi-crystal structure formed by the bouncing atoms is related to the sequence of bounces not to the sequence of time intervals between the bounces --- the latter can be different, see Fig.~\ref{LRstructure}. In the experiment, the time intervals which are very small can be disrupted due to imperfections of the motion of the mirrors which can result in defects in the Fibonacci quasi-crystal.

Long time stability of our phenomenon resulting from the coupling of the system to the subspace complementary to the resonant subspace requires further investigation but we expect that the considered quasi-crystal is a prethermal state.

To conclude, quasi-crystal structures can emerge in the time domain due to spontaneous breaking of discrete time translation symmetry of the time-periodic many-body Hamiltonian. They can be realized in ultra-cold atoms bouncing between oscillating atom mirrors if atoms are loaded to a resonant classical orbit. The latter can be done if an atomic Bose-Einstein condensate is prepared in a trap located at a classical turning point of a resonant trajectory and afterwards the trapping potential is turned off at a proper moment of time \cite{Giergiel2018a} --- the mirrors can be realized by two blue-detuned repulsive light sheets formed by focusing laser beams with cylindrical lenses. It results in a quantum state where all atoms occupy a single localized Wannier-like wavepacket that evolves along a resonant orbit. For sufficiently strong attractive interactions between atoms, the localized atomic wavepacket will perform evolution with a quasi-crystal structure in time and will not decay. In contrast, for negligible interactions, atoms will tunnel to other localized wavepackets evolving along the orbit what indicates decay of the quasi-crystal. 

We are grateful to Peter Hannaford for his valuable comments.
Support of the National Science Centre, Poland via Projects No.~2016/20/W/ST4/00314 (K.G.), QuantERA programme No.~2017/25/Z/ST2/03027 (A.K.) and  No. 2016/21/B/ST2/01095 (K.S.) is acknowledged. 

%%%%%%%%%%%%%%%%%%%%%%%%%%%%%%%%%%%%%%%%%%%%%%%%%%%%%%%%%%%%%%%%%%%%%%%%%%%%%%%%%%%%%%%
\newpage
\section{Supplemental material}

In this Supplemental Material we describe details of the description of the many-body system of atoms bouncing resonantly between two orthogonal oscillating atom mirrors in the presence of the gravitational force. We begin with a single-particle problem and then generalize the approach to the many-body case. We also discuss shortly experimental implementation and address the problem of stability of the quasi-crystal structures in time.

\subsection{Single-particle problem}

We start with a short introduction to the Floquet formalism \cite{Shirley1965} and then switch to the description of a single-particle system which we are interested in.  We show how to obtain an effective Floquet Hamiltonian that describes resonant dynamics of the system.

\subsubsection{Floquet states} 
\label{secfloquet}

Consider the following time-dependent Schr\"odinger equation
\bea
i \frac{\partial}{\partial t} \psi(x,t)=H(t)\;\psi(x,t)
\label{seq1}
\eea
with Hamiltonian having a discrete time translation symmetry $H(t+T)=H(t)$. According to the Floquet theorem, solution of the equation (\ref{seq1}) may be written as a linear combination of the functions of the form
\bea
\psi_k(x,t)=e^{-i \varepsilon_k t} \phi_k(x,t),
\label{sol1}
\eea
where $\phi_k(x,t+T)=\phi_k(x,t)$ are time-periodic with the same period $T$ as the Hamiltonian \cite{Shirley1965}. Substituting the solution (\ref{sol1}) into the Schr\"odinger equation (\ref{seq1}) we obtain 
\bea
{\cal H}\;\phi_k(x,t)= \varepsilon_k \phi_k(x,t),
\label{seq2}
\eea
where ${\cal H}=H(t)-i  \frac{\partial}{\partial t}$ is termed the Floquet Hamiltonian and $\phi_k(x,t)$ are so-called Floquet eigenstates. Since the function $\phi_k(x,t) e^{i \frac{2\pi n}{T} t}$, where $n$ is integer, is also a solution of the eigenequation (\ref{seq2}) corresponding to the eigenvalue $\varepsilon_k+n\frac{2\pi}{T}$, the quasi-energy spectrum is periodic with the period $\frac{2\pi}{T}$ and in the description of a system it is actually sufficient to restrict to a fragment of the spectrum, i.e. to a single Floquet zone of the width of $\frac{2\pi}{T}$.

\subsubsection{Single-particle Hamiltonian}

Let us consider an atom bouncing between two orthogonal oscillating atom mirrors which form the angle $\pi/4$ with the gravitational force vector. The Hamiltonian, in the standard gravitational units \cite{Giergiel2018a} but with the gravitational acceleration $g\rightarrow g/\sqrt{2}$, reads
\bea
H&=&\frac{p_x^2+p_y^2}{2}+x+y+F_x\left(x+\frac{\lambda_x}{\omega^2} \cos(\omega t+\Delta \phi)\right)
\cr &&+F_y\left(y+\frac{\lambda_y}{\omega^2}\cos(\omega t)\right),
\label{hlab}
\eea
where $F_x(x)$ and $F_y(y)$ describe the mirrors, i.e., the profile of the reflecting potentials along $x$ and $y$ directions, respectively. The mirrors oscillate harmonically with the frequency $\omega$ around $x=0$ and $y=0$ with the amplitudes $\lambda_x/\omega^2$ and $\lambda_y/\omega^2$. Description of the system is more convenient if we switch from the laboratory frame to the frame oscillating with the mirrors. In the classical case it can be done by means of the canonical transformation,
\bea
x'=x+\frac{\lambda_x}{\omega^2}\cos(\omega t+\Delta\phi), && y'=y+\frac{\lambda_y}{\omega^2}\cos(\omega t), \cr
p_x'=p_x-\frac{\lambda_x}{\omega}\sin(\omega t+\Delta\phi), && p_y'=p_y-\frac{\lambda_y}{\omega}\sin(\omega t),
\cr &&
\eea
while in the quantum case by the corresponding unitary transformation, i.e. $U_{y}=e^{iy\frac{\lambda_y}{\omega}\sin\omega t}e^{ip_y\frac{\lambda_y}{\omega^2}\cos\omega t}$ and a similar one for the motion along $x$. The resulting Hamiltonian is the following
\bea
H&=&\frac{p_x'^2+p_y'^2}{2}+x'+y'+\lambda_x x' \cos(\omega t+\Delta \phi) \cr
&&+\lambda_y y'\cos(\omega t)+F_x(x')+F_y(y').
\label{hp}
\eea
We assume that the mirrors can be modeled by hard wall potentials located at $x'=0$ and $y'=0$ and therefore we may drop the $F_x(x')$ and $F_y(y')$ in (\ref{hp}) keeping in mind that motion of a particle takes place for $x'\ge 0$ and $y'\ge 0$. In the following we also drop primes and the final single-particle Hamiltonian reads
\be
H=\frac{p_x^2+p_y^2}{2}+x+y+\lambda_x x \cos(\omega t+\Delta \phi)+\lambda_y y\cos(\omega t).
\label{h}
\ee

\subsubsection{Description of resonant dynamics}

Let us start with the classical mechanics.
The single-particle problem described by the Hamiltonian (\ref{h}) consists of independent motion along $x$ and $y$ directions. We are interested in a resonance driving and in order to describe resonant dynamics we perform canonical transformation to the so-called action-angle variables of the unperturbed problem (i.e. when $\lambda_x=\lambda_y=0$) \cite{Lichtenberg1992}. In these new canonically conjugate variables, the unperturbed Hamiltonian depends on the new momenta (the actions $I_x$ and $I_y$) only,
\bea
H_0(I_x,I_y)&=&\frac{p_x^2+p_y^2}{2}+x+y
\cr 
&=&\frac{(3\pi)^{2/3}}{2}\left(I_x^{2/3}+I_y^{2/3}\right).
\eea
If $\lambda_x=\lambda_y=0$, the actions are constant of motion ($I_x,I_y=$const) and the corresponding position variables (the angles $\theta_x$ and $\theta_y$) evolve linearly in time, i.e. $\theta_{x,y}=\Omega_{x,y}t+\theta_{x,y}(0)$ where 
\be
\Omega_x(I_x)=\frac{dH_0(I_x,I_y)}{dI_x}, \quad \Omega_y(I_y)=\frac{dH_0(I_x,I_y)}{dI_y},
\ee
are frequencies of an unperturbed periodic bouncing of a particle on the static mirrors.
The total Hamiltonian (\ref{h}) in the action-angle variables takes the form \cite{Buchleitner2002}
\bea
H&=&H_0(I_x,I_y)+\lambda_x\cos(\omega t+\Delta \phi)\sum_nh_n(I_x)e^{in\theta_x} \cr
&& +\lambda_y \cos(\omega t)\sum_nh_n(I_y)e^{in\theta_y},
\eea
where $h_0(I_{x,y})=\left(\frac{\pi I_{x,y}}{\sqrt{3}}\right)^{2/3}$ and $h_n(I_{x,y})=\frac{(-1)^{n+1}}{n^2}\left(\frac{3I_{x,y}}{\pi^2}\right)^{2/3}$ if $n\ne 0$.

The resonant driving of a particle corresponds to the conditions
\be
s_x\Omega_x(I_{x0})=\omega, \quad s_y\Omega_{y} (I_{y0})=\omega,
\ee
where $s_x$ and $s_y$ are integers and  $I_{x0}$ and $I_{y0}$ are resonant values of the actions.
In order to obtain an effective Hamiltonian that describes motion of a particle close to a resonant orbit we apply the classical secular approximation \cite{Lichtenberg1992,Buchleitner2002}. First, we extend the phase space of the system by the time $t$ variable and its canonically conjugate momentum $p_t=-H$ which play a role of additional coordinates. The Hamiltonian in the extended phase space ${\cal H}=H+p_{t}$ is the classical analogue of the quantum Floquet Hamiltonian, where $p_t \to -i \frac{\partial}{\partial t}$. Next, we perform a canonical transformation to the frame moving along a resonant orbit,
\bea
\Theta_{x}&=&\theta_{x}-\frac{\omega}{s_{x}}t, \\
\Theta_{y}&=&\theta_{y}-\frac{\omega}{s_{y}}t, \\
P_{t}&=&p_{t}+\frac{\omega I_x}{s_x}+\frac{\omega I_y}{s_y},
\label{moveframe}
\eea
which results in 
\bea
{\cal H}&=&H_0(I_x,I_y)-\frac{\omega I_x}{s_x}-\frac{\omega I_y}{s_y}+P_{t}
\cr &&+\lambda_x\cos(\omega t+\Delta \phi)\sum_nh_n(I_x)e^{in\Theta_x}e^{in\omega t/s_x} \cr
&& +\lambda_y \cos(\omega t)\sum_nh_n(I_y)e^{in\Theta_y}e^{in\omega t/s_y},
\eea
and carry out averaging over time keeping all dynamical variables fixed. The latter is allowed because in the moving frame (\ref{moveframe}) both the actions and the angles and $P_t$ are slowly varying quantities if we are close to the resonant orbit (i.e. $P_{x}=I_{x}-I_{x0}\approx 0$ and $P_{y}=I_{y}-I_{y0}\approx 0$) and if the time-dependent perturbation is weak, i.e.
\bea
\frac{d\Theta_{x}}{dt}&=&\Omega_x(I_{x})-\frac{\omega}{s_x}+{\cal O}(\lambda_x)\approx 0, \quad {\rm for}\;I_x\approx I_{x0}, \cr
\frac{d\Theta_{y}}{dt}&=&\Omega_y(I_{y})-\frac{\omega}{s_y}+{\cal O}(\lambda_y)\approx 0, \quad {\rm for}\;I_y\approx I_{y0}. \cr &&
\eea
The resulting effective Hamiltonian reads \cite{Lichtenberg1992,Buchleitner2002}
\bea
{\cal H}_{\rm eff}&=&\frac{P_x^2}{2m_{{\rm eff},x}}+V_x\cos(s_x\Theta_x+\Delta\phi) \cr
&&+\frac{P_y^2}{2m_{{\rm eff},y}}+V_y\cos(s_y\Theta_y)+P_t,
\label{hsec}
\eea
where $V_{x}=\lambda_{x}h_{-s_{x}}(I_{x0})$, $m_{{\rm eff},x}^{-1}=\frac{d^2H_0(I_{x0},I_{y0})}{dI_{x0}^2}$ and similar expressions for $V_y$ and  $m_{{\rm eff},y}$.

The Hamiltonian (\ref{hsec}) has been obtained within the classical approach. In order to switch to the quantum description one has to quantize (\ref{hsec}), i.e. $P_{x,y}=-i\frac{\partial}{\partial \Theta_{x,y}}$ and $P_t=-i\frac{\partial}{\partial t}$. The other option is to apply the quantum version of the secular approximation from the very beginning \cite{Berman1977} that leads to the same result if we choose $I_{x0}\gg 1$ and $I_{y0}\gg 1$ \cite{Giergiel2018a}. The secular Hamiltonian (\ref{hsec}) is time-independent which implies that $P_t=$constant and can be dropped. Actually in the quantum description, due to the time-periodicity of the system, eigenvalues $k\omega$ of $P_t$ are quantized  (i.e. $k$ is integer \cite{Buchleitner2002}) that makes the quasi-energy spectrum to repeat itself with the period $\omega$ as expected, see Sec.~\ref{secfloquet}. In the following we consider Floquet eigenstates corresponding to $k=0$.

If we focus on a resonance where $s_{x,y}\gg 1$, the Hamiltonian (\ref{hsec}) corresponds to a solid state problem of an electron moving in a two-dimensional space crystal. We will be considering the first energy band of the quantized version of (\ref{hsec}) and therefore the description of a resonantly driven particle can still be simplified. Indeed, superposing the Bloch wave eigenfunctions of (\ref{hsec}) corresponding to the first energy band we can construct Wannier functions $W_{\vect i=(i_x,i_y)}(\Theta_x,\Theta_y)=w_{i_x}(\Theta_x)w_{i_y}(\Theta_y)$ localized in different sites of the periodic potential in (\ref{hsec}). These Wannier functions in the laboratory frame appear as localized wavepackets moving along a classical resonant orbit with the period $T=s_xs_y2\pi/\omega$, i.e. $W_{\vect i}(x,y,t)=w_{i_x}(x,t)w_{i_y}(y,t)$ where $w_{i_x}(x,t)$ is periodic with the period $s_x2\pi/\omega$ and $w_{i_y}(y,t)$ is periodic with the period $s_y2\pi/\omega$. In the basis of the Wannier functions, i.e. when we restrict to wavefunctions of the form $\psi=\sum_{\vect i}a_{\vect i}W_{\vect i}$, the Hamiltonian (\ref{hsec}) reads \cite{Sacha2015,sacha15a}
\be
{\cal H}_{\rm eff}\approx-\frac{1}{2}\sum_{\langle \vect{i}, \vect{j}\rangle}J_{\vect i \vect j}\;a_{\vect i}^* a_{\vect j},
\label{hsectb}
\ee
where the sum runs over nearest neighbour sites of the potential in (\ref{hsec}) and 
\be
J_{\vect i \vect j}=-2\la W_{\vect i}|{\cal H}_{\rm eff}|W_{\vect j}\ra,
\ee
stand for amplitudes of tunneling of a particle between neighbouring sites.

Equation~(\ref{hsectb}) is a standard tight-binding model and it indicates that a resonantly driven particle is equivalent to a solid state problem if we use the basis of localized wavepackets $W_{\vect i}(x,y,t)$ which are evolving periodically along a resonant classical trajectory \cite{sacha15a}.

\subsection{Many-body problem}
\subsubsection{Many-body Hamiltonian}

In the present section we would like to switch from the single-particle problem to many ultra-cold atoms which are bosons and which are bouncing between two oscillating orthogonal mirrors. We focus on a resonant driving and restrict to the Hilbert subspace which is spanned by the localized Wannier wavepackets $W_{\vect i}(x,y,t)$ introduced in the previous section. In other words we consider the subspace spanned by the Fock states $|n_{(1,1)},\dots,n_{(s_x,s_y)}\ra$, where $n_{(i_x,i_y)}$ denotes number of bosons occupying a Wannier wavepacket $W_{\vect i=(i_x,i_y)}$. Restricting to this subspace and expanding the bosonic field operator in the Wannier basis,
\be
\hat \psi(x,y,t)\approx\sum_{\vect i}W_{\vect i}(x,y,t)\;\hat a_{\vect i},
\label{hatpsi}
\ee 
where $\hat a_{\vect i}$ are the standard bosonic anihilation operators, and substituting (\ref{hatpsi}) to the many-body Floquet Hamiltonian we obtain a many-body version of the tight-binding model (\ref{hsectb}),
\bea
\hat{\cal H}&=&\frac{1}{T}\int\limits_0^{T}dt\int dxdy\;\hat\psi^\dagger\left[H(t) +\frac{g_{0}}{2}\hat\psi^\dagger\hat\psi-i\partial_t\right]\hat\psi, \cr
&\approx&
-\frac{1}{2}\sum_{\langle \vect{i}, \vect{j}\rangle}J_{\vect i \vect j}\;\hat a_{\vect i}^\dagger\hat a_{\vect j}+\frac12\sum_{\vect i,\vect j}U_{\vect i\vect j}\;\hat a_{\vect i}^\dagger\hat a_{\vect j}^\dagger\hat a_{\vect j}\hat a_{\vect i},
\label{bhh}
\eea
where $H(t)$ is given in (\ref{h}), $T=s_xs_y2\pi/\omega$ and $g_0$ is the parameter (proportional to the atomic s-wave scattering length) that characterizes the potential of contact interactions between atoms, $g_0\delta(x)\delta(y)$. The effective interaction coefficients in (\ref{bhh}) read
\be
U_{\vect i \vect j}=\frac{2g_0}{T}\int\limits_0^Tdt\int dxdy\;|W_{\vect i}|^2|W_{\vect j}|^2,
\ee
for $\vect i\ne \vect j$ and similar $U_{\vect i\vect i}$ but by factor two smaller \cite{sacha15a,Giergiel2018}. The Hamiltonian (\ref{bhh}) is the standard Bose-Hubbard Hamiltonian written in the time-periodic basis. Application of a time-dependent basis requires corrections to the Hamiltonian coming from the time derivative of the basis vectors \cite{Lacki2013}. Here, it is included automatically when we perform the action of the Floquet operator, $H(t)-i\partial_t$, on the Wannier functions.

Typically the {\it on-site} interaction coefficients $U_{\vect i\vect i}$ are at least an order of magnitude larger than $U_{\vect i\vect j}$ for {\it long-range} interactions ($\vect i\ne\vect j$).  In the Letter we choose the parameters of the system so that all nearest neighbour tunneling amplitudes are the same, 
\be
J\equiv J_{\vect i\vect j}.
\ee

 The Hamiltonian (\ref{bhh}) is actually the Bose-Hubbard model which is a many-body counterpart of the single-particle tight-binding Hamiltonian (\ref{hsectb}). It is valid if the interaction energy per particle is much smaller than the gap $E_{\rm gap}$ between the first and the second energy bands of the single-particle problem (\ref{hsec}) \cite{sacha15a,Giergiel2018}. For example for the parameters used in Figs.~2-3 in the Letter, the maximal {\it on-site} interaction energy per particle $|U_{\vect i\vect i}|N/J\approx 80$, where $N$ is the total number of atoms, that is much smaller than the energy gap $E_{\rm gap}/J\approx 3000$.  
 
 \subsubsection{Mean-field approximation}
 
If the interaction between atoms are negligible (i.e. $g_0\rightarrow 0$), the ground state of the Bose-Hubbard Hamiltonian (\ref{bhh}) is a Bose-Einstein condensate where all atoms occupy a condensate wavefunction which is a uniform superposition of all Wannier functions
\be
\psi(x,y,t)\propto\sum_{\vect i}W_{\vect i}(x,y,t). 
\label{meansym}
\ee
That is, the many-body ground state reads 
\be
\Psi(x_1,y_1,\dots,x_N,y_N,t)=\prod\limits_{j=1}^N\psi(x_j,y_j,t).
\label{gsweak}
\ee
On the other hand, if interactions between atoms are attractive and sufficiently strong it is energetically favorable to group all atoms in a single localized Wannier wavepacket because it decreases the energy \cite{Albiez2005,Zin2008,Oles2010}. Then, we expect the ground state to be of the form $\Psi=\prod_{j=1}^NW_{\vect i}(x_j,y_j,t)$ where $\vect i=(i_x,i_y)$ is arbitrary. However, such a state cannot be a Floquet eigenstate of the system because it evolves with the period $T=s_xs_y2\pi/\omega$ while the discrete time translation symmetry of the Hamiltonian requires that all Floquet eigenstates must evolve with the period of the driving $2\pi/\omega$. In order to reconcile the energy and symmetry requirements, the ground state takes the form
\be
\Psi(x_1,y_1,\dots,x_N,y_N,t)\propto\sum\limits_{\vect i}\left(\prod\limits_{j=1}^NW_{\vect i}(x_j,y_j,t)\right),
\label{catp}
\ee 
which in the Fock states basis reads
\bea
|\Psi\ra&=&\frac{1}{\sqrt{s_xs_y}}\left(|N,0,\dots,0\ra+|0,N,0,\dots,0\ra+\dots\right. \cr
&&\left.+|0,\dots,0,N\ra\right).
\label{catf}
\eea
Such a ground state is actually a Schr\"odinger cat-like state, i.e. a superposition of macroscopic states, and it is extremely fragile to any perturbation. For example it is sufficient to measure position of a single atom and the Schr\"odinger cat state collapses to one of the states which form the superposition,
\be
\Psi\rightarrow \prod_{j=1}^NW_{\vect i}(x_j,y_j,t),
\label{colap}
\ee
in the Fock states basis it corresponds to
\be
|\Psi\ra\rightarrow|0,.\dots,0,N,0,0,\dots,0\ra.
\ee 
Which localized Wannier wavepacket $W_{\vect i}$ is chosen in (\ref{colap}) depends on the result of the measurement of the position of an atom. In other words this is an example of a spontaneous process which is responsible for spontaneous breaking of the discrete time translation symmetry of the Hamiltonian.

Note that both the ground state (\ref{gsweak}) of the weakly interacting system and a symmetry broken state (\ref{colap}) are Bose-Einstein condensates which can be described within the mean-field approximation. Thus, we may use the mean-field approach to describe the phenomenon we are after. In the mean-field description, the spontaneous time translation symmetry breaking will be indicated by appearance of a bifurcation where the symmetry-preserving ground state solution (\ref{gsweak}) looses its stability and new stable solutions are born which break the discrete time translation symmetry of the Hamiltonian and evolve with a period which is different from the driving period.

It is straightforward to obtain the mean-field equations. Indeed, the mean-field quasi-energy functional reads \cite{Sacha2015,Giergiel2018a}
\bea
E&=&\frac{1}{T}\int\limits_0^{T}dt\int dxdy\;\psi^*\left[H +\frac{g_{0}N}{2}|\psi|^2-i\partial_t\right]\psi, \cr
&\approx&
-\frac{1}{2}\sum_{\langle \vect{i}, \vect{j}\rangle}J_{\vect i \vect j}\;a_{\vect i}^*a_{\vect j}+\frac{N}{2}\sum_{\vect i,\vect j}U_{\vect i\vect j}\;|a_{\vect i}|^2 |a_{\vect j}|^2,
\label{mean}
\eea
where we have restricted to the resonant subspace with a condensate wavefunction $\psi(x,y,t)=\sum_{\vect i}a_{\vect i}W_{\vect i}(x,y,t)$. The ground state of (\ref{mean}) can be found by solving the corresponding Gross-Pitaevskii equation \cite{Pethick2002}. For negligible interactions, the ground state is of the form (\ref{meansym}) while for sufficiently strong attractive interactions there are $s_xs_y$ degenerate mean-field ground state solutions $\psi(x,y,t)\approx W_{\vect i}(x,y,t)$ corresponding to different values of $\vect i=(i_x,i_y)$.

The mean-field approach has been used in the Letter in order to obtain quantitative prediction for a range of the system parameters where time quasi-crystals form spontaneously. The time quasi-crystal states predicted by the mean-field approach lives forever. The predictions are valid in the limit when $N\rightarrow\infty$ but $g_0N=$const because then the corresponding symmetry preserving eigenstates of the quantum many-body Bose-Hubbard model (\ref{bhh}) are degenerate and their superpositions that form the symmetry-broken states do not decay \cite{Sacha2015}.

\subsection{Scenario for the experimental realization}

An oscillating atom mirror can be realized when an evanescent wave, created close to the surface of dielectric material, is modulated in time or by means of an oscillating blue-detuned repulsive light sheet formed by focusing a laser beam with cylindrical lens. The latter can be quite easily reconfigured into two orthogonal oscillating mirrors.

In order to demonstrate our finding in the laboratory one has to prepare a Bose-Einstein condensate of ultra-cold atoms in a trap at the position $(x,y)\approx(16,37)$ for the parameters used in Fig.~2 of the Letter, i.e. close to the classical turning point. Then, at the moment of time which is synchronized with the oscillations of the mirrors, the trapping potential has to be turned off and the atomic cloud starts falling on the mirrors. If the orientation of the mirrors is adjusted so that the gravitational force points along a fragment of the trajectory that connects the points $(x,y)\approx(16,37)$ and $(x,y)\approx(0,15)$, cf. Fig.~2 of the Letter, the atomic cloud falling on the mirrors is already moving along the proper resonant orbit. It means that all atoms are prepared in a single Wannier-like wavepacket $W_{\vect i}(\vect r,t)$ of the lower band. If the interactions between atoms are weak, one will observe tunneling of atoms to other Wannier-like wavepackets. However, if the interactions are sufficiently strong, atoms do not tunnel and the stable time quasi-crystal structure emerges in the course of the time evolution.  This strategy is similar to the strategy proposed for the experiments on discrete time crystals, see \cite{Giergiel2018a} for details. 

\subsection{Discussion of stability of periodically driven systems with classical non-linear resonances}

A periodically driven system can be heated by absorbing unlimited amount of energy like in the case of a resonantly driven harmonic oscillator. However, we deal with a system which, in the single-particle case, possesses non-linear classical resonances. The latter are well known in the field of dynamical systems or in quantum chaos \cite{Lichtenberg1992,Buchleitner2002}. In such systems a period of unperturbed motion of a particle depends on its energy and therefore a particle is not able to absorb continuously the energy: when the energy of a particle increases, its period changes, the system goes out of the resonance and the transfer of the energy stops. Such single particle resonantly driven systems are effectively described by secular Hamiltonians which are time-independent in the frame moving along the classical resonant orbit and which describe resonant elliptical islands created in the phase space \cite{Lichtenberg1992,Buchleitner2002}, see Eq~(\ref{hsec}). When these effective Hamiltonians are valid, a particle cannot be heated. The validity of an effective Hamiltonian can easily be monitored by looking at the exact structure of the classical phase space: if chaotic layers are absent, the effective Hamiltonian captures the exact dynamics of a particle \cite{Giergiel2018a}. The robustness of the resonantly driven systems against heating is related to stable resonant elliptical islands in the classical phase space.

In the many-body case, the interaction between particles introduces an additional perturbation. However, we consider the interaction energy per particle which is orders of magnitude smaller than the energy gap between the first and second resonant quasi-energy bands --- lack of the coupling of these bands is the most critical condition for the validity of the effective many-body Hamiltonian Eq.~(\ref{bhh}). Such a situation is similar to the standard tight-binding approximation commonly used in solid state physics and in ultra-cold atoms. That is, a single-band approximation is valid if the interaction energy per particle is much smaller than the energy gap between the bands. The effective description of many-body resonant behavior of atoms bouncing on an oscillating mirror was tested in a different context in Refs.~\cite{Sacha2015,Giergiel2018a}. There, it has been shown that the mean-field description of a Bose system within the full Gross-Pitaevskii equation is identical to the mean-field results obtained within the effective Hamiltonian approach.

%%%%%%%%%%%%%%%%%%%%%%%%%%%%%%%%%%%%%%%%%%%%%%%%%%%%%%%%%%%%%%%%%%%%%%%%%%%%%%%%%%%%%%%

%\bibliography{ref_11_2018_quasi} 

%%%%%%%%%%%%%%%%%%%%%%%%%%%%%%%%%%%%%%%%%%%%%%%%%%%%%%%%%%%%%%%%%%%%%%%%%%%%%%%%%%%%%%%

\end{document}